\documentclass[prd,twocolumn,nofootinbib,aps,tightenlines,preprintnumbers,superscriptaddress]{revtex4-1}

\usepackage{amsmath}

\usepackage[dvipsnames, usenames]{xcolor}
\definecolor{linkcolor}{rgb}{0.0,0.3,0.5}
\usepackage[unicode, colorlinks=true, linkcolor=linkcolor, citecolor=linkcolor, filecolor=linkcolor,urlcolor=linkcolor, pdfusetitle]{hyperref}

\usepackage{amsfonts}
\usepackage{graphicx}
\usepackage{dcolumn}
\usepackage{microtype}
\usepackage{bm}
\usepackage{color}
\usepackage{siunitx}
\usepackage[text={7in,9in},centering]{geometry}
\usepackage{aas_macros}
\newcommand{\code}[1]{\texttt{#1}}

\def\be{\begin{equation}}
\def\ee{\end{equation}}
\def\bea{\begin{eqnarray}}
\def\eea{\end{eqnarray}}

\begin{document}
\title{Gravitational-wave detection rates for compact binaries formed in isolation:\\LIGO/Virgo O3 and beyond}

\author{Vishal Baibhav}
\email{vbaibha1@jhu.edu}
\affiliation{Department of Physics and Astronomy, Johns Hopkins University, 3400 N. Charles Street, Baltimore, MD 21218, USA}

\author{Emanuele Berti}
\affiliation{Department of Physics and Astronomy, Johns Hopkins University, 3400 N. Charles Street, Baltimore, MD 21218, USA}

\author{Davide Gerosa}
\affiliation{School of Physics and Astronomy and Institute for Gravitational Wave Astronomy, University of Birmingham, \\Birmingham, B15 2TT, UK}

\author{Michela Mapelli}
\affiliation{Dipartimento di Fisica e Astronomia `G. Galilei', University of Padova, Vicolo dell'Osservatorio 3, I--35122, Padova, Italy}
\affiliation{INFN, Sezione di Padova, Via Marzolo 8, I--35131, Padova, Italy}
\affiliation{INAF-Osservatorio Astronomico di Padova, Vicolo dell'Osservatorio 5, I--35122, Padova, Italy}
\affiliation{Institut f\"ur  Astro- und Teilchenphysik, Universit\"at Innsbruck, Technikerstrasse 25/8, A--6020, Innsbruck, Austria}

\author{Nicola Giacobbo}
\affiliation{Dipartimento di Fisica e Astronomia `G. Galilei', University of Padova, Vicolo dell'Osservatorio 3, I--35122, Padova, Italy}
\affiliation{INFN, Sezione di Padova, Via Marzolo 8, I--35131, Padova, Italy}
\affiliation{INAF-Osservatorio Astronomico di Padova, Vicolo dell'Osservatorio 5, I--35122, Padova, Italy}

\author{Yann Bouffanais}
\affiliation{Dipartimento di Fisica e Astronomia `G. Galilei', University of Padova, Vicolo dell'Osservatorio 3, I--35122, Padova, Italy}
\affiliation{INFN, Sezione di Padova, Via Marzolo 8, I--35131, Padova, Italy}

\author{Ugo N. Di~Carlo}
\affiliation{INFN, Sezione di Padova, Via Marzolo 8, I--35131, Padova, Italy}
\affiliation{Dipartimento di Scienza e Alta Tecnologia, University of Insubria, Via Valleggio 11, I--22100, Como, Italy}

\date{\today}

\begin{abstract}
Using simulations performed with the population synthesis code \code{MOBSE}, we compute the merger rate densities and detection rates of compact binary mergers formed in isolation for second- and third-generation gravitational-wave detectors. We estimate how rates are affected by uncertainties on key stellar-physics parameters, namely common envelope evolution and natal kicks. We estimate how future upgrades will increase the size of the available catalog of merger events, and we discuss features of the merger rate density that will become accessible with third-generation detectors.
\end{abstract}

\maketitle

\section{Introduction}

The detection of gravitational waves (GWs) from $10$ binary black holes (BBHs) and a binary neutron star (BNS) in the first two LIGO/Virgo observing runs~\cite{LIGOScientific:2018mvr}, and the subsequent detections of  numerous compact binary candidates in the third observing run, naturally lead to the question: how do these binaries form, and what is the physics that drives their evolution?

Advanced LIGO (AdLIGO) is expected to reach design sensitivity in the near future, the so-called A$+$ upgrade to current detectors was already approved for funding, and further upgrades (A$++$ and Voyager) are expected in the near future~\cite{Aasi:2013wya,Adhikari:2013kya,Miller:2014kma,Voyager,Evans:2016mbw}. The GW community is also planning future, ``third-generation'' (3G) facilities, such as the Einstein Telescope (ET)~\cite{ETweb,Punturo:2010zz} and Cosmic Explorer (CE)~\cite{Evans:2016mbw}, which will extend the observable horizon to the very early Universe.

As GW detectors improve and the number of detections grows, we will gather information about the environments in which compact binaries form, and 
constrain the physical parameters that drive their evolution. Future GW detectors will measure compact binary parameters (such as masses and spins) within few per cent accuracy~\cite{Vitale:2016icu}, reconstructing fine details of distribution of these observables. They will observe sources up to redshifts as large as $z\sim 10^2$~\cite{Hall:2019xmm}, allowing us to study how the merger rate density evolves with redshift, and ultimately to constrain astrophysical models~\cite{Fishbach:2018edt,Vitale:2018yhm,LIGOScientific:2018jsj}. The large number of detections that comes with increased sensitivity will also reduce statistical errors on the parameters that describe compact binary populations to few per cent with $\sim 10^3$ observations ~\cite{Barrett:2017fcw}. 

Compact-object binaries could form either in the field~\cite{Belczynski:2001uc,Hurley:2002rf} or through dynamical interactions in young~\cite{Banerjee:2009hs,Ziosi:2014sra,Mapelli:2016vca}, nuclear~\cite{Antonini:2016gqe,Hoang:2017fvh} or globular clusters~\cite{Rodriguez:2016kxx,Askar:2016jwt}. In this paper we present updated detection rates, and a roadmap of our prospects for constraining the astrophysics of compact binaries in the near future. We study how detection rates for binaries formed in isolation (``field binaries'') will evolve with future improvements of GW detectors, with the goal to understand if and when characteristic features of the astrophysical populations will become visible.

The plan of the paper is as follows. In Sec.~\ref{sec:pops} we present our astrophysical populations based on the \code{MOBSE} population-synthesis code \citep{mapelli2017,Giacobbo:2017qhh}.
In Sec.~\ref{sec:MarRates} we investigate how uncertainties in binary evolution affect the evolution of the merger rate density, and what new generation of detectors can tell us about this evolution. In Sec.~\ref{sec:DetRates} we compute detection rates for each of the six models we consider and for different detector sensitivities.  In Sec.~\ref{sec:conclusions} we summarize our findings and out line directions for future work. Appendix~\ref{sec:DetRateCalculations} gives details on how detection rates are computed from the \code{MOBSE} simulations.  Throughout the paper we use the standard cosmological parameters determined by the Planck Collaboration~\cite{Ade:2015xua}. We assume that a source is detected if the single-detector signal-to-noise ratio (SNR) $\rho$ is such that $\rho>8$.

\section{Astrophysical populations}
\label{sec:pops}

We use simulations performed with the population-synthesis code \code{MOBSE} \cite{Giacobbo:2017qhh}. \code{MOBSE} is an upgrade of the \code{BSE} code \cite{Hurley:2002rf} which includes up-to-date prescriptions for the evolution of massive stars. The treatment of stellar winds accounts for the stellar metallicity and luminosity dependence of the mass loss. Compact objects are produced via different channels, including core-collapse, electron-capture and (pulsational) pair instability supernovae (SNe).

In our simulations, the primary star's mass $m_{\mathrm{1}}$ is distributed according to the Kroupa mass function \cite{Kroupa:2000iv}
\be
\mathcal{F}(m_1) \propto m_1^{-2.3} \qquad {\rm with} \;\; m_1 \in [5-150]M_\odot\,,
\ee
while the mass ratio $q=m_2/m_1$ scales like~\cite{Sana:2012px} 
\be
\mathcal{F}(q) \propto q^{-0.1} \qquad {\rm with} \;\; q \in [0.1-1]\,.
\ee
The orbital period $P$ is drawn from
\be
\mathcal{F}({\mathcal P}) \propto  {\mathcal P}^{-0.55} \quad {\rm with} \;\; {\mathcal P} = \mathrm{log_{10}}\left(\frac{P}{\mathrm{day}}\right) \in [0.15-5.5]
\ee
and the eccentricity $e$ follows the distribution~\cite{Sana:2012px}
\be
\mathcal{F}(e) \propto e^{-0.42} \qquad \mathrm{with}\;\; 0\leq e < 1\,.
\ee

Among the many physical processes involved in the formation of compact binaries that can merge within a Hubble time, the so called common-envelope phase is believed to be critical~\cite{Mapelli:2018uds,Mandel:2018hfr}. When a star in a binary system overfills its Roche lobe, it starts transferring mass, and eventually forms a common envelope that engulfs the companion. The common envelope does not corotate with the stars or their cores, and this leads to a drag force. As a result, the stars spiral in and transfer their orbital energy to the envelope. The system will survive only if the energy transferred is sufficient to eject the envelope~\cite{Ivanova:2013db,Ivanova:2012vx,Dominik:2012kk}.
  The efficiency of this mechanism constitutes a main uncertainty in  compact-binary formation modelling.

Another important source of uncertainty are natal kicks.
If a compact object forms from a supernova explosion, it is expected to receive a birth kick because of asymmetric mass ejection. A non-zero kick (the so-called Blaauw kick~\cite{1961BAN....15..265B}) is expected even in the unlikely case where mass loss is symmetric, but the compact object is part of a binary system. This natal kick can disrupt the binary or substantially modify its orbit. Kicks set the fraction of stellar binaries which are unbound by the SN explosion and, consequently, play a major role in determining GW detection rates \cite{Belczynski:2001uc,mapelli2017,Gerosa:2018wbw}.

\begin{table}[t]
  \caption{Catalog of \code{MOBSE} models considered in this study.}
\label{tab:MobseModels}
\setlength\tabcolsep{15 pt}
\begin{tabular}{rrr}
\hline
Model &  $\sigma_{\rm{CCSN}}$  &    $\alpha$  \\
\hline
$\alpha1$  & $265\,{\rm km/s}$  & $1$ \\
$\alpha3$  & $265\,{\rm km/s}$ &  $3$ \\
$\alpha5$  & $265\,{\rm km/s}$ &  $5$ \\
CC15$\,\alpha1$ &  $15\,{\rm km/s}$ & $1$ \\ 
CC15$\,\alpha3$ &  $15\,{\rm km/s}$ &  $3$ \\ 
CC15$\,\alpha5$ &  $15\,{\rm km/s}$ &  $5$ \\ 
\hline
\end{tabular}
\end{table}

As described by \citet{Giacobbo:2018etu} and summarized in Table~\ref{tab:MobseModels}, we consider six representative populations of merging binaries, aiming at bracketing the uncertainties in the physics of both common envelope and natal kicks. These two parameters might be the first to be constrained with GW data (see e.g. \cite{OShaughnessy:2017eks,Barrett:2017fcw}).

The common envelope phase is treated using the so-called $\alpha\lambda$ formalism~\cite{Webbink:1984ti,Ivanova:2012vx}, where $\alpha$ quantifies the efficiency of energy transfer to the envelope and $\lambda$ represents the binding energy of the envelope. In this work we consider $\alpha$ as a free parameter, while $\lambda$ depends on the stellar type~\cite{Dewi:2000nq} and it is computed by using the prescriptions derived in Ref.~\cite{2014A&A...563A..83C}.
Kicks are extracted from a Maxwellian distribution with root-mean-square speed (rms) $\sigma_{\rm CCSN}$ for core-collapse SNe that produce neutron stars.\footnote{Neutron stars can also form through electron-capture SNe, which are less energetic, faster and do not develop large asymmetries. This is generally expected to lead to small kicks, and therefore we assume $\sigma_{\rm ECSN}=15 \,{\rm km/s}$ \citep{giacobbo2019}.} For black holes, we reduce the kick velocity $v_{\rm BH}$ by taking into account fallback: $v_{\rm BH} = (1 - f_{\rm fb})v_{\rm NS}$, where $v_{\rm NS}$ is the natal kick for neutron stars and $f_{\rm fb}$ parametrizes the amount of fallback on the proto-compact object~\cite{Fryer:2011cx}. 

Models CC15 produce natal kicks $\leq{}100$ km s$^{-1}$, and therefore they are in tension with the proper motions of the fastest single Galactic neutron stars~\cite{Hobbs:2005yx}. These models were chosen because they give a local merger rate density of binary neutron stars consistent with the one inferred from GW170817~\cite{TheLIGOScientific:2017qsa}, without requiring exotic assumptions about common envelope.

\code{MOBSE} predicts the NS masses from $1.1$ to $2 M_\odot$ where light (heavy) NSs are preferred during BNS (NSBH) mergers. On the other hand, NSBH mergers favor low BH masses ($<15 M_\odot$) while BBH mergers could have BHs as heavy as $45 M_\odot$ with most binaries having mass ratios close to unity~\cite{Giacobbo:2018etu}.

\section{Merger rate densities}
\label{sec:MarRates}
\begin{figure}
\centering
  \includegraphics[width=0.99\columnwidth]{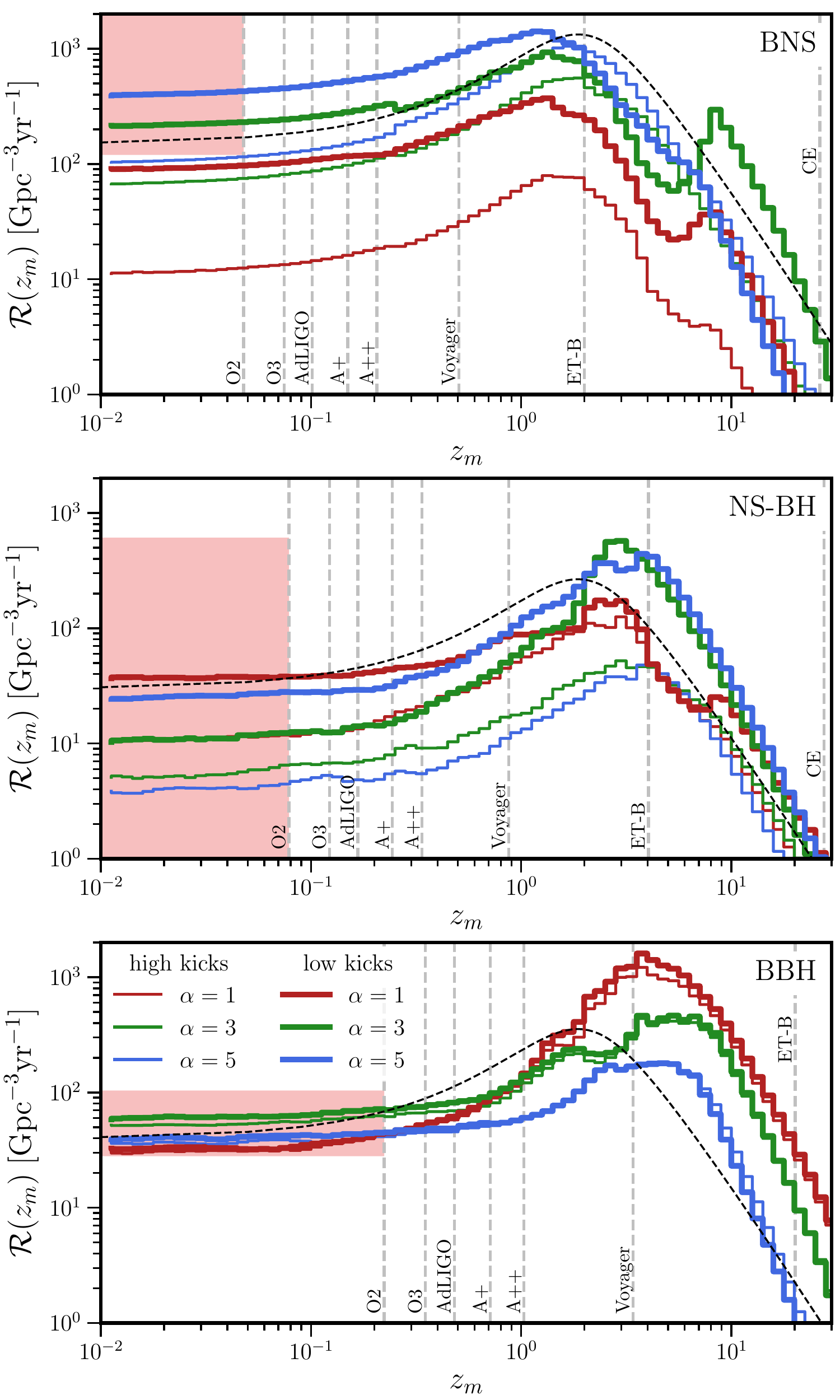}
  \caption{Merger rate density ${\mathcal R}(z_m)$ for the models listed in Table~\ref{tab:MobseModels}. Here ``low kicks'' corresponds to $\sigma_{\rm CCSN}=15$~km/s, while ``high kicks'' corresponds to $\sigma_{\rm CCSN}=265$ km/s. Black dashed lines are proportional to the star formation rate. Vertical dashed gray lines correspond to the horizon obtained by assuming BNSs of mass $(1.4+1.4)\,M_\odot$, NSBHs of mass $(1.4+5)\,M_\odot$,  and BBHs of mass $(10+10)\,M_\odot$ (see~\cite{Chen:2017wpg} for a discussion).  For BBHs, the CE horizon $z=77$ is so large that it lies to the right of the x-axis range in the figure. The red shaded region shows the allowed ranges for the merger rate densities based on O1 and O2 observations with their ``power law'' model for BBHs and ``uniform mass" model for BNSs obtained using the PyCBC pipeline).}  \label{fig:MergerRateDensity} \end{figure}

\begin{figure*}[htp]
\centering
  \includegraphics[width=0.99\textwidth]{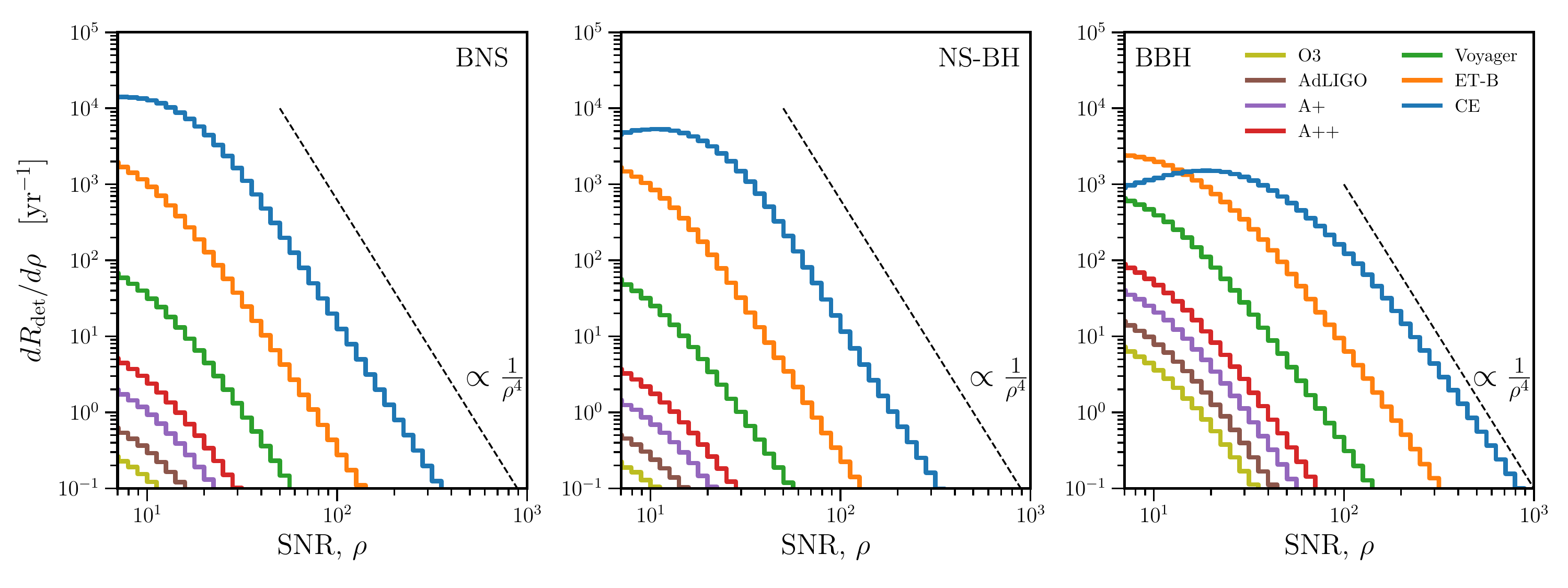}
  \caption{SNR distribution for the low-kick $\alpha=5$ model and different detectors. Here $R_{\rm det}$ is the number of detections per year for the given detector, as defined in Eq.~(\ref{eq:Rdet1}).}
\label{fig:SNRdistribution}
\end{figure*}

The merger rate density ${\mathcal R}(z_m)$ as a function of merger redshift $z_m$ tracks the distribution of merging binaries across cosmic time, and it depends on two factors:
\begin{itemize}
\item[(i)] the rate of binary formation at a given redshift $z_f$, and
\item[(ii)] the distribution of time delays $t_{\rm delay}$ between the formation of the parent stars in the binary and the merger of their compact object remnants.
\end{itemize}
In turn, binary formation at $z_f$ depends on the star formation rate and the metallicity, both of which evolve over time. The time delay distribution is sensitive to the physics that drives binary evolution (see e.g.~\cite{Mandel:2015qlu,Rodriguez:2016kxx,DiCarlo:2019pmf}). 

In Fig.~\ref{fig:MergerRateDensity} we plot the evolution of the merger rate density for the six \code{MOBSE} models considered in this study. The low-redshift behavior is often parametrized as a power law: ${\mathcal R}(z)\approx {\mathcal R}_0 (1+z)^{\lambda_0}$~\cite{LIGOScientific:2018jsj,Fishbach:2018edt}, where ${\mathcal R}_0$ is the local merger rate density and $\lambda_0$ is a model-dependent parameter that describes its evolution. The parameter $\lambda_0$ can be used to infer astrophysical information. The star formation rate is well approximated by $\lambda_0\simeq 2.4$ for $0.1<z<1$~\cite{Fishbach:2018edt}. Therefore, an observed $\lambda_0<2.4$ would imply that mergers peaked before the peak of star formation, which is only possible if compact-object binary formation is high at low metallicities and if the time delays are short enough~\cite{Fishbach:2018edt}. Current detectors can only investigate the evolution of the merger rate at low redshift, but in the near future we will be able to trace the redshift evolution of the merger rate density.

Figure~\ref{fig:MergerRateDensity} shows that the BNS rate density follows quite closely the star formation rate, with a peak at slightly lower redshift (because of the short but finite time delays). Current observations favor models with low kicks and large $\alpha$: as shown by the red shaded region in the top panel of Fig.~\ref{fig:MergerRateDensity}, only low-kick models with $\alpha=3$ and $\alpha=5$ can explain the high local merger rates resulting from the detection of GW170817 \citep{mapelli2018,Giacobbo:2018etu}.  Most BNS formation models have weak dependence on metallicity. Quite interestingly, some of them show a bimodal distribution, with a dip at $z_m\approx5.6$ and a second peak at $z_m\approx9$.  Indeed, the efficiency in forming merging BNS has a minimum at intermediate metallicity $Z \sim 0.1 Z_{\odot}$ (see e.g. Fig.~14 of \cite{Giacobbo:2018etu}). Stars at intermediate metallicities tend to develop larger radii, and this leads to the formation of wide BNS systems that either do not merge in a Hubble time, or are easily disrupted by a SN explosion (because of their large orbital separation).  However, not all models that show a dip in the merger efficiency lead to a bimodal merger rate density. Since most detectors are not sensitive to binaries from such large redshifts, 3G detectors are needed to observe this behavior in the early Universe.

By contrast, BBH production is very efficient at low metallicities because of the impact of metallicity on stellar radii and evolutionary stages. At solar metallicity massive stars become Wolf-Rayet stars quite rapidly, after leaving the giant branch, because of stellar wind efficiency. Wolf-Rayet stars have small radii ($1-2~R_\odot$); thus, it is highly unlikely that such stars enter common envelope. Without common envelope, the binary star evolves into a BBH with a large orbital separation, which will not be able to merge within a Hubble time. In contrast, metal-poor massive stars can retain a large fraction of their hydrogen envelope and avoid the Wolf-Rayet stage, increasing the probability of undergoing mass transfer and entering common envelope.
The rate density peaks at $z\gtrsim 2$, earlier than the peak of star formation, and the merger rate density at small redshifts is not as steep as the star formation rate (i.e., it has $\lambda_0<2.4$). We should soon be able to verify this trend with current detectors.

\section{Detection rates}
\label{sec:DetRates}

\begin{figure*}
\centering
  \includegraphics[width=0.99\textwidth]{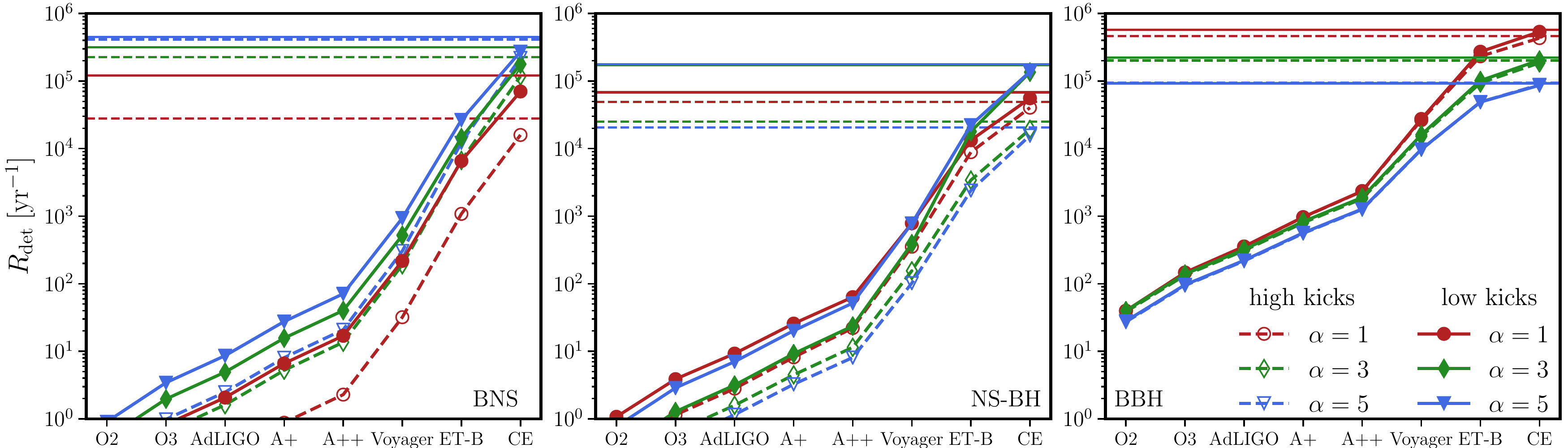}
  \caption{Detection rates of BBHs, NSBHs, and BNSs for second- and third-generation detectors. Here ``low kicks'' corresponds to $\sigma_{\rm CCSN}=15$~km/s, while ``high kicks'' corresponds to $\sigma_{\rm CCSN}=265$~km/s. Horizontal lines represent all events in the universe, as would be seen by a perfect (noiseless) detector.
} 
\label{fig:totalrate}
\end{figure*}

\begin{table*}
  \caption{Minimum and maximum detection rates (${\rm yr}^{-1}$) across all models.}
\label{tab:detRates}
\begin{center}
 \setlength\tabcolsep{20 pt}
\begin{tabular}{lccc}
\hline
\multicolumn{1}{c}{Detector} & \multicolumn{1}{c}{BNS} & \multicolumn{1}{c}{NSBH} & \multicolumn{1}{c}{BBH} \\
\hline
O2 & \num{0.028}\,{\rm -}\,\num{0.91} & \num{0.12}\,{\rm -}\,\num{1.1} & \num{27}\,{\rm -}\,\num{40} \\
O3 & \num{0.11}\,{\rm -}\,\num{3.4} & \num{0.46}\,{\rm -}\,\num{3.9} & \num{94}\,{\rm -}\,\num{1.5e+2} \\
AdLIGO & \num{0.27}\,{\rm -}\,\num{8.6} & \num{1.2}\,{\rm -}\,\num{9.3} & \num{2.2e+2}\,{\rm -}\,\num{3.6e+2} \\
A+ & \num{0.88}\,{\rm -}\,\num{28} & \num{3.2}\,{\rm -}\,\num{26} & \num{5.6e+2}\,{\rm -}\,\num{9.7e+2} \\
A++ & \num{2.3}\,{\rm -}\,\num{71} & \num{8.1}\,{\rm -}\,\num{63} & \num{1.3e+3}\,{\rm -}\,\num{2.4e+3} \\
Voyager & \num{32}\,{\rm -}\,\num{9.4e+2} & \num{1.0e+2}\,{\rm -}\,\num{7.8e+2} & \num{9.7e+3}\,{\rm -}\,\num{2.7e+4} \\
ET-B & \num{1.1e+3}\,{\rm -}\,\num{2.7e+4} & \num{2.4e+3}\,{\rm -}\,\num{2.2e+4} & \num{4.9e+4}\,{\rm -}\,\num{2.7e+5} \\
CE & \num{1.6e+4}\,{\rm -}\,\num{2.7e+5} & \num{1.6e+4}\,{\rm -}\,\num{1.4e+5} & \num{8.6e+4}\,{\rm -}\,\num{5.4e+5} \\
Noiseless & \num{2.8e+4}\,{\rm -}\,\num{4.5e+5} & \num{2.0e+4}\,{\rm -}\,\num{1.8e+5} & \num{9.2e+4}\,{\rm -}\,\num{5.7e+5} \\
\hline
    \end{tabular}
  \end{center}
\end{table*}

To study how detection rates will benefit from detector improvements, here we will consider noise power spectral densities for  the AdLIGO design sensitivity noise~\cite{Aasi:2013wya}; planned upgrades to existing LIGO detectors (A+, A++ and Voyager~\cite{Miller:2014kma,Adhikari:2013kya,Voyager}); and 3G detectors, including CE~\cite{Evans:2016mbw} and the Einstein Telescope (more specifically, ET-B~\cite{ETweb}). We approximate the detector noise for the O2 and O3 runs by rescaling the AdLIGO noise curve in such a way that the resulting BNS range is $90$~Mpc~\cite{LIGOScientific:2018mvr} and $140$~Mpc~\cite{DetectorStatusGWOSC}, respectively.
In Fig.~\ref{fig:SNRdistribution} we plot the distribution of signal-to-noise ratios (SNRs) for these detectors using the low-kick model with $\alpha=5$. Most of the binaries with very large SNRs come from local Universe, so their distribution scales like $1/ \rho{}^4$~\cite{Schutz:2011tw}.\footnote{In the local Universe, the total number of binaries within luminosity distance $D_*$ is $N(D<D_*)\propto D_*^3$, or equivalently  $N(\rho>\rho_*)\ \propto \rho_*^{-3}$, so the SNR probability distribution scales like $N(\rho_*)=\frac{dN(\rho>\rho_*)}{d\rho_*}\propto \rho_*^{-4}$.}
Since CE (and, for BBHs, also ET) will see past the peak of the merger rate density (cf. Fig.~\ref{fig:MergerRateDensity}), the maximum detection redshift is not controlled by the detector capabilities, but by the physics that governs the merger rate density ${\mathcal R}(z_m)$. Figure~\ref{fig:totalrate} shows the detection rates, $R_{\rm det}$ for different astrophysical models and different detectors, comparing them with the intrinsic merger rate in the Universe that would correspond to an ideal, {\em noiseless} detector (see Appendix~\ref{sec:DetRateCalculations} for details of the detection-rate calculations).  According to our models, AdLIGO at design sensitivity could see $220-360$ BBH, up to $9$ NSBH and $9$ BNS mergers per year. Upgrading AdLIGO detectors to a configuration like A+ would increase the detection rates by a factor of $3$. With 3G detectors, BBH rates would increase by up to $2$--$3$ orders of magnitude, while NSBH and BNS detection rates would increase by up to $3$--$4$ orders of magnitude. CE would see at least $92\%$ of all BBH mergers in the Universe, compared to the $0.06$--$0.24\%$ seen by AdLIGO at design sensitivity. Current-generation detectors like AdLIGO have low BNS and NSBH detection rates, detecting only $10^{-5}$ ($\sim 10^{-4}$) of all BNS (NSBH) mergers in the Universe. By contrast, CE will see more than $50\%$ ($\sim 75\%$) of all BNS (NSBH) mergers.

 It is also clear from Fig.~\ref{fig:totalrate} that $\alpha$ and $\sigma_{\rm CCSN}$ can affect detection rates of all compact binary systems by up to an order of magnitude. In particular, BBH and BNS rates are affected in different ways by the common-envelope efficiency parameter $\alpha$: lower values of $\alpha$ yield smaller rates for BNSs and larger rates for BBHs.
This can be understood as follows. BBHs form from massive stars that can develop very large radii during their evolution, and therefore enter the common envelope phase with a wide orbital separation. If $\alpha>1$, the envelope will be ejected easily while the binary is still widely separated, and the outcome will be a wide binary that is unlikely to merge in a Hubble time~\cite{Giacobbo:2018etu}.
In contrast, BNSs form from smaller stars, and the orbital separation at the beginning of the common envelope phase is smaller. Therefore high values of $\alpha$ lead to the formation of a close binary that can merge in a Hubble time, while small values of $\alpha$ cause a premature merger of the system.

Low kicks (CC$15\alpha1$,  CC$15\alpha3$,  CC$15\alpha5$) lead to higher detections rates for BNS and NSBH mergers, because strong kicks are efficient at disrupting these binaries. On the other hand, most BBH progenitors undergo direct collapse in the models presented here: nearly all of the star's mass falls back onto the compact object, and kicks are suppressed. For this reason,  BBH detection rates are nearly insensitive to natal kicks.\footnote{BBH merger rates are found to strongly depend on SN kicks if fallback is suppressed~\citep{mapelli2017,Wysocki:2017isg,Gerosa:2018wbw}.} 

Local NSBH merger rates for low-kick models are larger than high-kick models by a factor of 3--10. If we assume low (high) SN kicks, the NSBH merger rate increases (decreases) with $\alpha$. This is because large SN kicks tend to unbind the binary. If the natal kick is high, a small value of $\alpha{}$ increases the probability that the system merges, because  if $\alpha$ is small the system's semi-major axis shrinks considerably during CE, after the first supernova. Thus, if the kick is high a small value of $\alpha{}$ increases the NSBH merger rate. In contrast, if the kick is low, a small value of $\alpha{}$ might trigger the premature merger of the binary, before the second compact object has formed. Thus, if the kick is low, the highest NSBH merger rate is achieved for a rather large value of $\alpha{}$, as already explained in \cite{mapelli2018}.

We list minimum and maximum rates across all models in Table~\ref{tab:detRates}. 

\section{Conclusions}
\label{sec:conclusions}

We studied the detection rates and redshift evolution of BNS, NSBH and BBH merger rate densities. The redshift distribution of the merger rates contains important clues about the physics that drives the evolution of these compact objects (see also the companion papers \cite{mapelli2017,mapelli2018,mapelli2019}). The merger rate history of compact-object binaries is obtained by convolving their formation history with the time-delay distribution. The formation rate depends on both star formation rate and metallicity. The formation of BNSs depends only mildly on metallicity, and therefore their formation across cosmic time follows quite closely the star formation rate (but it is shifted to slightly lower redshifts, because of finite delay times).  Therefore for BNSs we expect $\lambda_0\gtrsim2.4$, i.e. the merger rate peak occurs after, but very close to the peak of star formation.  Current detectors have small BNS horizons, so they will mainly see binaries that formed in the local Universe, where metallicity is high, but 3G detectors should allow us to observe large-redshift BNSs and to verify this prediction.
In contrast, BBH production (and, marginally, NSBH production) is very efficient at low metallicities. Most BBHs form at $z \gtrsim 2$, before the peak of star formation, and their merger rate density evolves slowly compared to BNSs: most BBHs and NSBHs formed before the peak of star formation, yielding $\lambda_0<2.4$. 
Only CE (and, in the case of BBHs, ET) will allow us to see beyond the merger rate peak of compact object binaries. 

We also investigated how these rates are affected by common-envelope efficiency and natal kicks, considering both second- and third-generation detectors. We found that a lower common envelope efficiency leads to smaller BNS detection rates, and larger BBH detection rates. This is because lower efficiency causes a longer inspiral of the stellar cores, leading to BNS progenitors that merge prematurely, before they can collapse into a neutron star. By contrast, BBH progenitors are much larger, and their orbits are wider compared to BNS progenitors. Natal kick assumptions affects only BNS and NSBH mergers in our models: high kicks can more easily disrupt binaries  and {\em usually} lead to lower detection rates. On the other hand, BBH kicks are suppressed because of the large amount of material that falls back onto the compact object after the supernova explosion.

\begin{figure}[t]
\centering
  \includegraphics[width=\columnwidth]{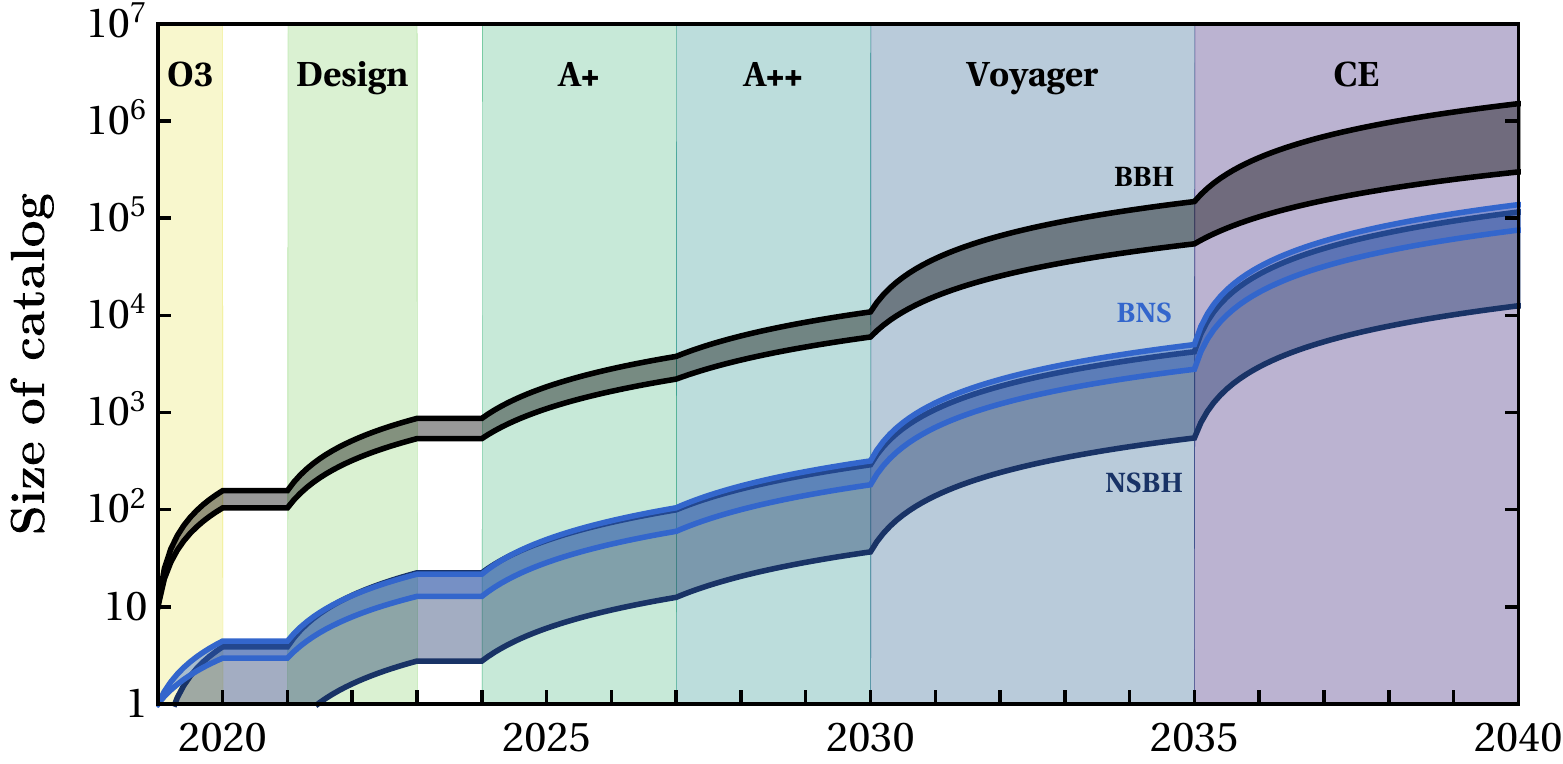}
  \caption{Growth of catalog size as detectors improve for models in agreement with current observations. The timeline for different detectors and their upgrades is estimated following Refs.~\cite{LIGOScientific:2019vkc,InstrumentWhitePaper2017,InstrumentWhitePaper2018}. We assume an optimistic duty cycle of 100\%, which is compatible with expectations for future observations with multiple detectors.}  \label{fig:catalogsize} \end{figure}

In Fig.~\ref{fig:catalogsize} we plot the growth of the GW catalog size as detectors improve, based on the rate calculations of Fig.~\ref{fig:totalrate}. We assume 1 year of observations for O3, which started in 2019. The observing run O4 for AdLIGO at design sensitivity is expected to start in 2021, and it should last for $\sim 2$~years, followed by 1 year of commissioning period for upgrades to A+ (which is currently targeted to be operational by 2024~\cite{InstrumentWhitePaper2018}). We assume the operational time for A+ to be 6~years~\cite{InstrumentWhitePaper2017}, with further upgrades to ``A++'' in 2027. By the beginning of the 2030s, when new detectors -- Voyager in the existing LIGO facilities, and CE/ET in separate facilities -- may start operations, we could have a GW catalog of up to $10^4$ events. In Fig.~\ref{fig:catalogsize} we assume a 5-year observation period before Voyager is superseded by CE.

As the detectors improve, the rapid growth of the GW catalog should allow us to place stringent constraints on the population parameters that influence the final stages of the lives of massive stars.

\section*{Acknowledgments}
MM and YB  acknowledge financial support by the European Research Council for the ERC Consolidator grant DEMOBLACK, under contract no. 770017. %
EB and VB are supported by NSF Grant No. PHY-1841464, NSF Grant No. AST-1841358, NSF-XSEDE Grant No. PHY-090003, and NASA ATP Grant No. 17-ATP17-0225.  
This work has received funding from the European Union’s Horizon 2020 research and innovation programme under the Marie Sk\l{}odowska-Curie grant agreement No. 690904.  The authors would like to acknowledge networking support by the COST Action GWverse CA16104.
Computational work was performed on the University of Birmingham's BlueBEAR cluster and at the Maryland Advanced Research Computing Center (MARCC).

\appendix
\section{Detection rate calculations }
\label{sec:DetRateCalculations}

The detection rate is given by \cite{Dominik:2014yma,Belczynski:2015tba}
\bea \label{eq:Rdet1}
R_{\rm det} = \int_0^{t_0} p_{\rm det}{\mathcal R} (z_m)   \frac{dV_c}{dt_m} \frac{dt_m}{dt_{\rm det}} dt_m ,
\eea
where $t_0$ is the age of universe and $p_{\rm det}$ is the probability of detecting a given binary, defined in Eq.~(\ref{eq:pdet}) below. The factor
${dt_m}/{dt_{\rm det}} = {1}/{(1+z_m)}$ accounts for the different clock rates 
at merger and at the detector. The source-frame merger rate density at redshift $z_m$ is
\bea \label{eq:mergerrate1}
{\mathcal R} (z_m) &\equiv& \frac{dN}{dV_c dt_m} = \int_0^{t_m}   {\rm sfr}(z_f) 
\frac{dN} {dt_m dM_f} dt_f,
\eea
where the star-formation rate is ${\rm sfr}(z_f)\equiv\frac{dM_f}{dV_c dt_f} $. The second term in the integrand accounts for the number of binaries per unit star-forming mass that form at $t_f$ and merge at $t_m$. Here, we have marginalized over the distribution of component masses and time delays. We can rewrite Eq.~\eqref{eq:Rdet1} (after switching the order of the integrals over $t_f$ and $t_m$)  as
\bea \label{eq:Rdet2}
R_{\rm det} &=& \int_0^{t_0} {\rm sfr}(z_f)\frac{d\quad}{dM_f}\left( \int_{t_f}^{t_0} 
\frac{dN} {dt_m}  \frac{p_{\rm det}(z_m)}{1+z_m}   \frac{dV_c}{dt_m}  dt_m \right) dt_f,\nonumber\\
&=& \int_0^{t_0} {\rm sfr}(z_f)\frac{d\quad}{dM_f}\left( \sum  \frac{p_{\rm det}(z_m)}{1+z_m}  \frac{dV_c}{dz_m} \frac{dz_m}{dt_m} \right) dt_f.
\eea
In the second line above, we converted the integral over a distribution to a Monte-Carlo sum, 
\be
\int \frac{dN}{dt_m} f(t_m) dt_m\to\sum_{i} f(t^{i}_{m})\;.
\ee
In practice, the term in parentheses is evaluated  by Monte Carlo integrations, where the samples $t^{i}_{m}$ are generated from the distribution ${dN}/{dt_m}$.
The comoving volume element $dV_{\rm c}/dz$ is given by

\begin{equation}
\frac{dV_c}{dz}(z) =4 \pi \frac{c}{H_0} \frac{D_{\rm c}^2 }{ E(z)},
\end{equation}
where $E(z)$ is the function that describes the evolution of Hubble parameter, i.e. $H(z)=H_0 E(z)$, and
$D_{\rm c}$ is comoving distance~\cite{Hogg:1999ad}. The factor of $4\pi$ takes into account the angular integration over the sky.

 In practice, at a given metallicity $Z_f$, \code{MOBSE} starts with a given total mass $M_{\rm sim}$ and outputs a distribution of binaries. For each set of free parameters in Table~\ref{tab:MobseModels}, we have $12$ simulations of $10^7$ binaries each, with metallicities $Z= 0.01$--$1\,Z_\odot $. We simulate a set of compact-object binaries formed at different times $t_f$  inside bins of $\Delta t_f=10{\rm ~Myr}$. At the time of formation $t_f$, we assume that the metallicity is given by 
\bea\label{eq:Zevolution}
\log{\frac{Z(z_f)}{Z_{\odot}}}=
\begin{cases}
-0.19\ z_f, & z_f \leq 1.5\\
-0.22\ z_f, & z_f > 1.5\,,
\end{cases}
\eea
i.e. we follow the metallicity evolution of Ref.~\cite{2012ApJ...755...89R}, but we rescale it so that $Z(0)=Z_\odot$. Each formation time bin is assigned one the $12$ metallicities according to Eq.~\eqref{eq:Zevolution}. However, since the \code{MOBSE} simulation started with total binary mass, $M_{\rm sim}$, we need to rescale this mass according to the star formation in that particular time bin. We have adopted the following fit for star formation rate \cite{Madau:2014bja}:
\begin{equation}\label{eq:SFR}
{\rm sfr}(z) =  \frac{0.015(1+z)^{2.7}}{1 + [(1+z)/2.9]^{5.6}} \, M_{\odot} {\rm Mpc}^{-3} . 
\end{equation}
These binaries are then evolved in time until they merge at $t_m$. This produces a catalog of binaries that form at $t_f$ and merge at $z_m$. The integral in Eq.~\eqref{eq:Rdet2} can be now be written as
\be
R_{\rm det}=\sum_i ( s_i(t_f)\Delta t_f ) {\frac{p_{\rm det}}{ 1+z_m}} {\frac{dV_{\rm c}}{dz_m}} {\frac{dz_m}{dt_m}}  \,, 
\ee 
where all terms except the first are evaluated at the merger redshift $z_m$. The first term is the number density of binaries formed at redshift $z_f$,
\begin{equation}
s_i(z_f)\Delta t_f = f_{\rm bin} f_{\rm IMF} \frac{ {\rm sfr}(z_f)}{M_{\rm sim}(Z_f)} \Delta t_f \,.
\end{equation}
The factors $f_{\rm bin}=0.5$ and $f_{\rm IMF}=0.285$ take into account the fact that \code{MOBSE} only simulates binaries with primary
mass larger than $5 M_{\odot}$. 

Finally, a binary is assumed to be detected if it has the signal-to-noise ratio (SNR) $\rho=\rho_0 w>8$, where $\rho_0$ is the SNR assuming that the binary is optimally oriented and located in the sky, while $0\leq w \leq 1$ is the projection factor that depends on the binary's sky position and orientation. 
The optimal SNR is calculated as
\be \label{eq:defSNR}
\rho^2_0 = 4\int_0^\infty \frac{\tilde h^*(f) \tilde h(f)}{S_h(f)}df\,,
\ee
where $h(f)$ is the frequency-domain GW signal and $S_h(f)$ is the detector noise power spectral density~\cite{Moore:2014lga,Sathyaprakash:2009xs}. The horizon $z_h$ is the farthest redshift for which a binary with component masses $m_1$ and $m_2$ can be detected, i.e. $\rho_0(m_1,m_2,z_h)=8$.
The quantity $\rho_0$ determines the probability of detecting a binary that lies within the detector's horizon (i.e. $\rho_0>8$, or equivalently $z<z_h$):
\be
p_{\rm det}=\int_{8/\rho_0}^1 p(\omega) d\omega
\label{eq:pdet}
\ee
where $p(w)$ is the probability distribution function of $\omega$ \cite{Finn:1992xs}. Detection rates only depend on $p_{\rm det}$, hence $\rho_0$. We calculate the signal-to-noise ratio of BBH mergers using the waveform approximant \code{IMRPhenomD}, while for NSBH and BNS mergers we use \code{TaylorF2}. Since \code{MOBSE}
does not have any prescriptions to evolve the spins, we assume black holes and neutron stars to be non-spinning. Spins are expected to impact detection rates within a factor 1.5~\cite{Gerosa:2018wbw}, which should be added to the error budget of our estimates.

Note that in Fig~\ref{fig:SNRdistribution}, where we looked at the distribution of $\rho= \rho_0 w$, we sample $p(\omega)$ for each binary in the catalogs mentioned above and assign the SNR accordingly. 

\bibliography{detection_rate}
\end{document}